\documentclass[letterpaper]{jpconf}
\usepackage{graphicx}
\begin{document}
\title{Higher Charmonium}

\author{T.Barnes}

\address{Department of Physics and Astronomy, University of Tennessee,
Knoxville, TN 37996, USA\\
Physics Division, Oak Ridge National Laboratory,
Oak Ridge, TN 37831, USA}

\ead{tbarnes@utk.edu}

\begin{abstract}
This contribution briefly discusses some new theoretical results 
for the properties of charmonium, especially the ``higher charmonium" 
states above open-charm threshold. In particular
we discuss the spectrum of states, open-flavor strong decays, and 
the surprisingly large effects of virtual decay loops of charmed meson
pairs.
\end{abstract}

\section{Introduction}

Recently there have been several experimental reports 
of hadrons with remarkable and unexpected properties. 
These include the surprisingly low mass charm-strange mesons 
D$_{s0}^{*+}(2317)$ 
\cite{Aubert:2003fg}
and
D$_{s1}^{+}(2460)$ 
\cite{Besson:2003jp},
the charmonium or charmed meson molecule 
candidate X(3872)
\cite{Choi:2003ue},
and the pentaquark candidate $\theta(1540)$  
\cite{Nakano:2003qx,Kubarovsky:2003fi}.
These reports motivate careful, detailed studies of the physics 
underlying the resonances observed in these sectors, both to determine
what properties are expected for accessible states where the calculations are
incomplete, and also to determine ``what has gone wrong"  
with the existing models. In this contribution we  
abstract some predictions for properties 
of higher mass charmonium states 
(specifically the spectrum and strong decay widths)
from one such detailed study, which is currently in preparation \cite{BGS};
some closely related results for charmonium were published 
previously in a study of possible X(3872) assignments 
\cite{Barnes:2003vb}.

\section{Spectrum}

\begin{figure}[h]
\begin{center}
\includegraphics[width=28pc]{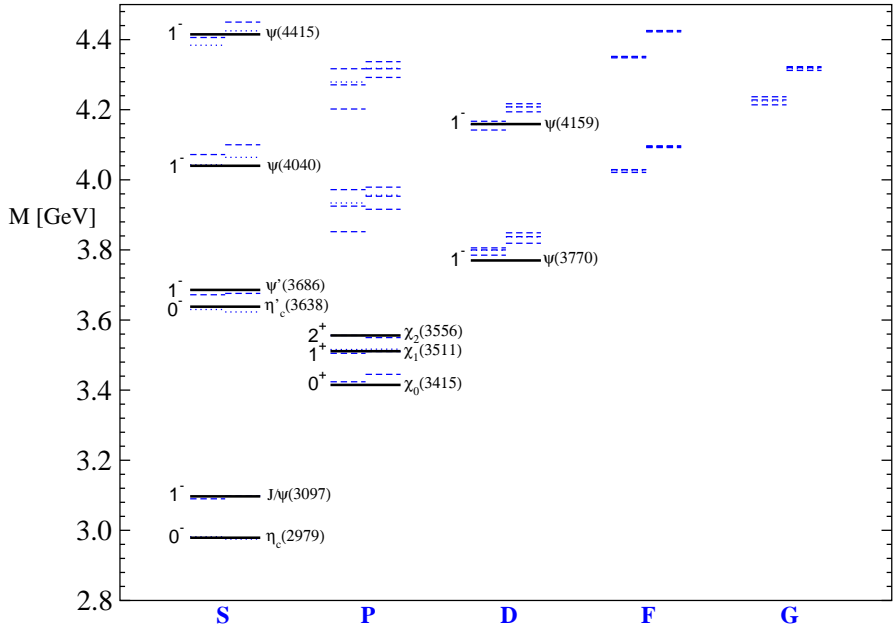}
\end{center}
\caption{
\label{spectrum}
The spectrum of states (to L=4) predicted by the 
nonrelativistic (left) 
and Godfrey-Isgur (right) potential models. 
The well-established experimental states used as input for the NR model
are also shown (solid).}
\end{figure}

It is well known that a simple potential model gives a remarkably good
description of the charmonium spectrum. The spectrum predicted by a model
of this type, including all multiplets with entries below 4.4 GeV, is 
given in Table~\ref{table_spectrum}. This ``minimal" model uses the 
nonrelativistic Schr\"odinger equation, with a zeroth-order potential 
consisting of the color Coulomb plus linear scalar confining terms and a
Gaussian-smeared OGE spin-spin hyperfine interaction. The OGE spin-orbit and 
tensor and scalar confinement spin-orbit terms are incorporated in first-order 
perturbation theory. The four parameters $\alpha_s$ (OGE strength), 
$b$ (string tension), $m_c$ (charm quark mass) and $1/\sigma$ 
(hyperfine smearing length) are determined by a fit to the well-established 
experimental states given in the table. 
The results are shown in Fig.\ref{spectrum}, together with the 
spectrum predicted by the 
relativized Godfrey-Isgur model~\cite{Godfrey:1985xj}. The spectra are
rather similar, although the NR model gives somewhat lower masses for 
higher-L states, largely due to the choice of a smaller string tension.
We note in passing that recent quenched LGT results for the spectrum of
charmonium \cite{Liao:2002rj}
are very similar to the predictions of these potential models. 

The generic features of the resulting spectrum, which are well known, 
are slowly decreasing radial excitation energy gaps with increasing N and L, 
and rapidly decreasing splittings within a multiplet
with increasing L (due to suppressed short-distance wavefunctions).
For sufficiently large L the multiplets invert, due to the increased 
importance of the inverted scalar spin-orbit term. This effect
however may be obscured by other mass shifts.

\begin{table}[h]
\caption{\label{table_spectrum}Spectrum of charmonium states in a 
nonrelativistic potential model (masses in MeV). 
The experimental states used as input are 
shown in brackets, and the resulting
parameters are $\alpha_s = 0.5461$, $b = 0.1425$~GeV$^2$,
$m_c = 1.4794$~GeV and $\sigma = 1.0946$~GeV. 
All multiplets with any state below 4400~MeV are shown.}
\begin{center}
\begin{tabular}{lllll}
\br
Multiplet 
& M$_{\rm S=1}^{\rm J=L+1}$
& M$_{\rm S=1}^{\rm J=L\phantom{+1}}  $
& M$_{\rm S=1}^{\rm J=L-1}$
& M$_{\rm S=0}^{\rm J=L\phantom{+1}}  $  
\\
\mr
input: \\
1S & [3097] & & & [2979] \\
2S & [3686] & & & [3638] \\
3S & [4040] & & &  \\
4S & [4415] & & &  \\
1P & [3556] & [3511] & [3415] &  \\
1D &  &  & [3770] &  \\
2D &  &  & [4159] &  \\
predictions: \\
1S& 3090 & & & 2982 \\
2S& 3672 & & & 3630 \\
3S& 4072 & & & 4043 \\
4S& 4406 & & & 4384 \\
1P& 3556 & 3505 & 3424 & 3516 \\
2P& 3972 & 3925 & 3852 & 3934 \\
3P& 4317 & 4271 & 4202 & 4279 \\
1D& 3806 & 3800 & 3785 & 3799 \\
2D& 4167 & 4158 & 4142 & 4158 \\
1F& 4021 & 4029 & 4029 & 4026 \\
2F& 4348 & 4352 & 4351 & 4350 \\
1G& 4214 & 4228 & 4237 & 4225 \\
1H& 4392 & 4410 & 4424 & 4407 \\
\br
\end{tabular}
\end{center}
\end{table}

\section{Strong Widths}

Strong width predictions for higher-mass charmonium states are of great 
importance, since they indicate which states should be narrow enough to
identify easily, and which modes should be favored by a given state.
The dominant strong decays (when energetically allowed) are 
open flavor decays, which in the valence approximation are due to the 
process $(c\bar c) \to (c \bar q )(q \bar c)$. (Here, $q=u,d,s$.) 
This strong decay mechanism is surprisingly poorly understood in terms of 
QCD degrees of freedom \cite{Ackleh:1996yt},
and is usually (for light hadrons) described using the phenomenological
``$^3$P$_0$ model" \cite{Micu:1968mk}. 
This model assumes that the new $q\bar q$ pair is produced 
with vacuum quantum numbers, with a universal dimensionless amplitude $\gamma$ that is
determined by the data. (Light meson decays suggest a value of 
$\gamma \approx 0.4$~\cite{Barnes:1996ff}.)
In charmonium an alternate model, in which the $q\bar q$ pair is produced
by a linear vector confining interaction, was introduced by the Cornell 
group \cite{Eichten:ag} 
and has been employed in recent studies by Eichten {\it et al}
\cite{Eichten:2004uh}.

Application of this model to charmonium gives predictions for all the
two-body open charm decay amplitudes and partial widths of every 
state, which is clearly a great deal of information. 
In view of the limited space, here we only quote the 
partial and total widths of the four known states above DD threshold,
which are the $\psi(3770), \psi(4040), \psi(4159)$ and $\psi(4415)$. 
These results are given in Table~\ref{table_decays}.

Little is known about these strong branching fractions experimentally,
which is unfortunate because a striking preference for certain modes is
predicted, such as DD$_1$ being the largest mode of the $\psi(4415)$. 
The weakness of $\psi(4040) \to {\rm DD}$ agrees
with experiment \cite{Goldhaber:1977qn}, but the exclusive branching 
fractions of the $\psi(4159)$ and $\psi(4415)$ have unfortunately not
been measured. These exclusive decays will hopefully be studied at CLEO-c 
and BES in the near future. We also note that the mode D$^*$D$^*$ is 
very interesting, in that it has three $1^{--}$ waves, $^1$P$_1$, 
$^5$P$_1$ and $^5$F$_1$, and the relative decay amplitudes depend 
strongly on the nature of the initial $c\bar c$ state. For S-wave 
$c\bar c$ mesons (presumably including the 
$\psi(4040)$ and $\psi(4415)$) the $^3$P$_0$ model predicts
the amplitude ratio $^5$P$_1 /  ^1$P$_1 = -2\sqrt{5}$ 
and $^5$F$_1 = 0$, whereas for the D-wave $\psi(4159)$ one finds
a dominant $^5$F$_1$ amplitude, and $^5$P$_1 /  ^1$P$_1 = -1/\sqrt{5}$. 
A measurement of these amplitude ratios
would provide a sensitive test of the decay model 
(assuming that the usual spectroscopic assignments are correct).

\begin{table}[h]
\caption{\label{table_decays} 
$^3$P$_0$ model predictions for the partial and total widths (MeV) 
of known charmonium states above DD threshold. This
assumes SHO wavefunctions with a width parameter $\beta = 0.5$~GeV,
a pair production strength  $\gamma = 0.4$, and the usual
spectroscopic assignments 
$\psi(3770) = 1^3$D$_1$, 
$\psi(4040) = 3^3$S$_1$, 
$\psi(4159) = 2^3$D$_1$ and 
$\psi(4415) = 4^3$S$_1$.}
\begin{center}
\begin{tabular}{ccccccccccl}
\br
State &  DD & DD$^*$ & D$^*$D$^*$ 
& D$_s$D$_s$ & D$_s$D$_s^*$  &  D$_s^*$D$_s^*$ 
& DD$_1$ & DD$_1'$ & DD$_2^*$ & $\Gamma_{tot}$ [expt.] \\
\mr
$\psi(3770)$ &  43. &      &      &      &     &  
&&&& 43. [$23.6 \pm 2.7$]\\
$\psi(4040)$ &  0.1 & 33.  & 33.  & 7.8  &     &  
&&&& 74. [$52\pm 10$]  \\
$\psi(4159)$ &  16. &  0.4 & 35.  & 8.0  & 14. &  
&&&& 74. [$78 \pm 20$]\\
$\psi(4415)$ &  0.4 &  2.3 & 16.  & 1.3  & 2.6 & 
0.7 & 31. & 1.0 & 23. 
&  78. [$43 \pm 15$] \\
\br
\end{tabular}
\end{center}
\end{table}

\section{Loop Effects}

At second order 
in the decay process
one finds contributions to the 
composition and properties of hadrons due to virtual hadron loops. 
In charmonium, the process
$(c\bar c) \to (c\bar n) (n\bar c) \to (c\bar c)$ 
describes the virtual transition of a ``bare" quark model $c\bar c$ meson 
into two open-charm mesons. 
The associated
energy shift has real and imaginary parts, which give respectively 
the mass shift due to two-meson continuum mixing and the 
decay rate. (The imaginary part is zero if the state is below threshold.)
The first-order 
correction to the $|c\bar c\rangle$
state is of the form $|(c\bar n) (n\bar c)\rangle$, 
which specifies the continuum components of the charmonium resonance. 

The size of these non-valence components and their effects on observables
are interesting topics, since they are neglected in the valence 
approximation to the quark model and in quenched lattice QCD. The possibility
that loop effects (mixing with the two-meson continuum) may be responsible 
for the anomalously low masses of the new D$_{s{\rm J}}$ states has been suggested 
by several groups
\cite{Barnes:2003dj,vanBeveren:2003kd,Hwang:2004cd},
and was the principal motivation for our study of loop effects.   

We have tested the importance of loop effects in charmonium by calculating the
mass shifts and state composition for several charmonium states that result 
from loops of S-wave $c\bar n$ and $c\bar s$ meson pairs 
(six channels), using the $^3$P$_0$ model with the same parameters used 
previously to describe decays \cite{BS}.
The results for all 1S, 1P and 2S $c\bar c$ states are shown in Table~\ref{table_loops}. 
\begin{table}[h]
\caption{\label{table_loops} 
The effect of virtual meson loops on charmonium states below DD threshold.
The mass shifts (MeV) due to each mixing channel and the residual 
$|c\bar c\rangle$ 
probability P$_{c\bar c}$ are shown.}
\begin{center}
\begin{tabular}{l|ccccccc|c}
\br
\multicolumn{1}{c|}{State}
&
\multicolumn{7}{c|}{Mass Shift by Channel (MeV)}
&
\\
&\quad DD &\quad  DD$^*$ &\quad D$^*$D$^*$
&\quad  D$_s$D$_s$ &\quad  D$_s$D$_s$$^*$ &\quad D$_s$$^*$D$_s$$^*$
& \quad Total \quad & $\;$ P$_{c\bar c}$ \quad \\
\hline
$|1^3{\rm S}_1 (J/\psi) \rangle $
&  $-30.$  & $-108.$ & $-173.$ &  $-17.$  & $- 60.$ & $- 97.$ &  $-485.$
&  0.65  
\\
$|1^1{\rm S}_0 (\eta_c) \rangle $  
&  $\phantom{-}0$  & $-149.$ & $-137.$ &  $\phantom{-}0$  & $- 84.$ & $- 78.$
&  $-447.$
&  0.71 
\\ 
\hline
$|1^3{\rm P}_2 (\chi_{c2}) \rangle $  
&  $-53.$  & $-137.$ & $-188.$ &  $-22.$  & $- 59.$ & $- 79.$ 
&  $-537.$
&  0.43  
\\ 
$|1^3{\rm P}_1 (\chi_{c1}) \rangle $  
&  $\phantom{-}0$  & $-165.$ & $-194.$
&  $\phantom{-}0$  & $- 66.$ & $- 85.$
&  $-511.$
&  0.46  
\\
$|1^3{\rm P}_0 (\chi_{c0}) \rangle $  
&  $-75.$  & $\phantom{-}0$ & $-255.$
&  $-28.$  & $\phantom{-}0$ & $-113.$
&  $-471.$
&  0.53  
\\
$|1^1{\rm P}_1 (h_c) \rangle $  
&  $\phantom{-}0$  & $-194.$ & $-169.$
&  $\phantom{-}0$  & $- 81.$ & $- 73.$
&  $-516.$
&  0.46  
\\
\hline
$|2^3{\rm S}_1 (\psi') \rangle $
&  $-36.$  & $-110.$ & $-165.$
&  $-11.$  & $- 36.$ & $- 56.$
&  $-413.$
&  0.45  
\\
$|2^1{\rm S}_0 (\eta_c') \rangle $  
&  $\phantom{-}0$  & $-154.$ & $-134.$
&  $\phantom{-}0$  & $- 52.$ & $- 46.$
&  $-386.$
&  0.58  
\\
\br
\end{tabular}
\end{center}
\end{table}
Evidently the mass shifts and continuum mixing predicted by the $^3$P$_0$ 
model are indeed quite large, and it is surprising {\it a priori} 
that $c\bar c$ potential 
models describe the spectrum as accurately as they do 
(recall Fig.\ref{spectrum}). 
Presumably this is 
because the large negative mass shift is similar for all the low-lying 
charmonium states, and can be approximated by a change in the 
charm quark mass. 
Note in this regard that the residual scatter of total mass shifts  
within each multiplet (mean variance $19.$~MeV)
in Table~\ref{table_loops} is much smaller than the mean shift ($-471.$~MeV). 
If the assumed initial ``bare" masses of the states within an N,L 
$c\bar c$ multiplet are set equal, as are the masses of the charmed mesons 
in the loops,
we actually find that the total mass shift is the same for every state 
within the multiplet.
In any case, we conclude that loop effects are quite large, 
and should certainly be incorporated in future studies of the effects of 
``unquenching the quark model". 

\section*{Acknowledgements}

This research was supported in part by
the U.S. National Science Foundation through grants 
NSF-PHY-0244786 
and
NSF-INT-0327497
at the
University of Tennessee,
and by
the U.S. Department of Energy under contract
DE-AC05-00OR22725 at
Oak Ridge National Laboratory (ORNL).

\end{document}